\documentclass[10pt,twocolumn,preprint]{sigplanconf}

\usepackage{graphicx}
\usepackage{epsfig}
\usepackage{amssymb}
\usepackage{times}
\usepackage{url}
\usepackage{xspace}
\usepackage{verbatim}
\usepackage{amsmath}
\usepackage{subfigure}
\usepackage{multirow}
\usepackage[usenames,dvipsnames]{color}
\usepackage{hyperref}
\usepackage{breakurl}

\usepackage[font=bf]{caption} 

\makeatletter
\def\@copyrightspace{\relax}
\makeatother

\newcommand{\aaron}[1]{}
\newcommand{\robert}[1]{}
\newcommand{\aditya}[1]{}
\newcommand{\ashok}[1]{}
\newcommand{\theo}[1]{}
\newcommand{\vyas}[1]{}
\newcommand{\maybenix}[1]{#1}
\newcommand{\maybenixB}[1]{#1}

\newcommand{\minisection}[1]{\smallskip \noindent{\bf #1.}}

\newcounter{linenum}

\newcommand{\Section}{\S}
\newcommand{\secref}[1]{{\S\ref{#1}}}

\newcommand{\figref}[1]{{Figure~\ref{#1}}}

\newcommand{\tabref}[1]{{Table~\ref{#1}}}

\newcommand{\compactcaption}[1]{\caption{#1}\vspace{-1em}}

\newenvironment{compactenumerate}
{
   \begin{enumerate}
   \vspace{-1ex}
   \setlength{\topsep}{0pt}
   \setlength{\itemsep}{0em}
   \setlength{\parskip}{0pt}
   \setlength{\parsep}{0pt}
}
{
   \vspace{-1ex}
   \end{enumerate}
}

\newenvironment{compactitem}
{
   \begin{itemize}
   \vspace{-1ex}
   \setlength{\topsep}{0pt}
   \setlength{\itemsep}{0em}
   \setlength{\parskip}{0pt}
   \setlength{\parsep}{0pt}
}
{
   \vspace{-1ex}
   \end{itemize}
}

\newcommand{\Mb}{{MBox}\xspace}

\newcommand{\mb}{{MBox}\xspace}
\newcommand{\mbs}{{MBoxes}\xspace}

\newcommand{\Name}{{Stratos}\xspace}

\newcommand{\VMb}{{{MB}}\xspace}
\newcommand{\VMbs}{{{MBs}}\xspace}

\newcommand{\downpaths}{\ensuremath{\mathit{downstream}}}

\newcommand{\ChainArray}{\ensuremath{\mathit{C}}}
\newcommand{\ChainArrayUnmang}{\ensuremath{\mathit{C}'}}
\newcommand{\chainposition}{\ensuremath{\mathit{j}}}
\newcommand{\numinchain}{\ensuremath{|\chaininstance|}}
\newcommand{\chaininstance}{\ensuremath{\mathit{c}}}

\newcommand{\MiddleboxSet}{\ensuremath{\mathit{M}}}
\newcommand{\middleboxinstance}{\ensuremath{\mathit{m}}}
\newcommand{\middleboxindex}{\ensuremath{\mathit{i}}}

\newcommand{\Volume}{\ensuremath{\mathit{V}}}
\newcommand{\Split}{\ensuremath{\mathit{f}}}
\newcommand{\Available}{\ensuremath{\mathit{b}}}
\newcommand{\Footprint}{\ensuremath{\mathit{Cost}}}

\newcommand{\Gain}{\ensuremath{\gamma}}
\newcommand{\nextto}{\ensuremath{\mathit{'}}}
\newcommand{\prevto}{\ensuremath{\mathit{'}}}
\newcommand{\rackinstance}{\ensuremath{\mathit{r}}}

\newcommand{\NoFlow}{HeavyWgt\xspace}
\newcommand{\MonitorVMs}{LocalView\xspace}
\newcommand{\UniformFlow}{UniformFlow\xspace}

\begin{document}


\setlength{\tabcolsep}{0.1cm}

\title{\Name: A Network-Aware Orchestration Layer for\\Virtual Middleboxes in
    Clouds} 

\authorinfo
{
    Aaron Gember$^{\ddag}$, 
    Anand Krishnamurthy$^{\ddag}$, 
    Saul St. John$^{\ddag}$, 
    Robert Grandl$^{\ddag}$,
    Xiaoyang Gao$^{\ddag}$,
    Ashok Anand$^{\dag}$,
    Theophilus Benson$^{\ast}$, 
    Vyas Sekar$^{\circ}$,
    Aditya Akella$^{\ddag}$ 
}
{
    $^{\ddag}$University of Wisconsin -- Madison, 
    $^{\dag}$Instart Logic,
    $^{\ast}$Duke University, 
    $^{\circ}$Stony Brook University
}

\maketitle

\begin{abstract}
 
Enterprises want their in-cloud services to leverage the performance and
security benefits that middleboxes offer in traditional deployments. 
 Such virtualized deployments create new opportunities (e.g.,
flexible scaling) as well as  new challenges (e.g., dynamics, multiplexing) for
middlebox management tasks such as service
 composition and provisioning.  Unfortunately, enterprises lack
systematic tools to efficiently  compose and provision in-the-cloud middleboxes
and thus fall short of achieving the benefits that cloud-based deployments can
offer. To this end,  we present the design and implementation of \Name, an
orchestration layer for virtual middleboxes.  \Name provides efficient and
correct composition in the presence of dynamic scaling via software-defined
networking mechanisms. It ensures efficient and scalable provisioning by combining
 middlebox-specific traffic engineering, placement, and horizontal scaling
strategies. We demonstrate the effectiveness of \Name  using an experimental
prototype testbed and large-scale simulations.



\end{abstract}

\section{Introduction}
\label{s:intro}

Surveys show that enterprises rely heavily on in-network middleboxes (\mbs)
such as load balancers, intrusion prevention systems, and WAN optimizers to
ensure application security and to improve performance~\cite{sherry2012making,
sekar2011middlebox}. As enterprises move their applications and services 
to the cloud, they would naturally like to realize these \mb-provided
performance and security benefits in the cloud as well. Recent 
  industry trends further confirm  this 
 transition  with an increasing number of virtual network
appliances~\cite{aryaka,silverpeak,suricata}, and in-the-cloud network
services~\cite{plumgrid,midokura,oneconvergence,embrane,aws}.




At a high level, virtualized \mb deployments create new challenges as well as
new opportunities for \mb\ {\em composition} (also referred to as service
chaining~\cite{quinn2013network}) and {\em provisioning} to meet desired
performance objectives. The proprietary and non-deterministic nature of \mbox
processing make these tasks hard even in traditional network
deployments~\cite{qazi2013simplefying,joseph2008policyaware}---the {\em
dynamic}, {\em virtualized}, and {\em multiplexed} nature of cloud deployments
compound the problem, leading to brittleness, inefficiency and poor scalability
(\secref{s:challenges}).

\smallskip
\noindent {\bf \mb Composition} Enterprises often need to chain multiple \mbs
together; e.g., traffic from a gateway must pass through a firewall, caching
proxy, and intrusion prevention system (IPS) before reaching an application
server. Today, such policy is implemented by physically wiring the topology.
Virtualization offers a new opportunity to break this coupling between the
policy and topology. At the same time,  the chaining
logic must now be implemented {\em via forwarding mechanisms}, which raises
unique challenges in the face of dynamic changes to \mb chains. In
particular, the stateful nature of \mbs coupled with the complexity of packet
{\em mangling}  operations they perform (e.g., NATs rewrite headers and proxies
terminate sessions)  makes it difficult to ensure forwarding
correctness and efficiency.

\smallskip
\noindent{\bf \mb Provisioning} Traditional \mb deployments are typically
overprovisioned or require drops in functionality in the presence of load;
e.g., an IDS may disable DPI capabilities under load~\cite{snort}.
Virtualized cloud deployments offer the ability to flexibly scale \mb
deployments as needs change. At the same time, the heterogeneity in \mb
processing, characteristics of \mb workloads, and multiplexed nature of cloud
deployments makes it challenging to address resource bottlenecks in an
efficient and scalable manner.
Furthermore, poor network placement or routing may  introduce network-level
effects that may adversely impact provisioning decisions by causing needless
scaling. 






While many \mb vendors are already making virtual \mbs readily available  to
enable enterprises to deploy in-cloud \mbs,  there's a dearth of systematic
tools to address the above composition and provisioning challenges.  To this
end, we design and implement, \Name, {\em a new network-aware orchestration
layer for virtual \mbs in clouds}.  

\Name provides a novel, flexible software-defined networking (SDN) solution
to the composition problem that leverages the virtualized nature of the
deployment. In contrast to prior SDN solutions that require expensive and
potentially inaccurate in-controller correlation or changes to
\mbs~\cite{qazi2013simplefying,fayazbakhsh2013flowtags}, \Name engineers a
simpler solution by marginally over-provisioning an \mb chain to explicitly
avoid potential steering ambiguity in the presence of mangling \mbs (\secref{s:composition}).

To ensure efficient and scalable provisioning, \Name employs a
scalable multi-level approach that carefully synthesizes ideas from traffic
engineering~\cite{heorhiadi2012new}, network-aware virtual machine
placement~\cite{benson2011cloudnaas, wood2007blackbox},
    and elastic compute scaling~\cite{rightscale} (\secref{s:provisioning}). As a first and light-weight step, it uses a flow distribution mechanism to address transient compute or network bottlenecks, without having to know the nature of the bottleneck itself. When this is insufficient, \Name locates persistent compute or network bottlenecks, and applies progressively heavier techniques via network-aware horizontal scaling and migration. The techniques ensure that the network footprint of \mb chains is low, and compute resource utilization is maximized meeting our efficiency goals.



We have implemented a fully featured \Name prototype ($\approx$12K lines of
Java code), including a forwarding controller written as a FloodLight
module~\cite{floodlight} and a stand-alone resource controller. We evaluate
this prototype in a 36 machine testbed using a variety of \mb chains and
synthetic request workloads. We also simulate \Name to understand its
properties at larger scale. We find that our composition mechanisms impose a
1ms overhead on the completion time per flow for each mangling \mb included in
a chain. By construction, \Name always maintains correct composition, whereas
state-of-the-art techniques have $\approx$19\% error rate in the presence of
mangling and dynamic provisioning~\cite{qazi2013simplefying}. Our provisioning
mechanisms satisfy application objectives using up to one-third fewer resources
and invoking up to one-third fewer heavy-weight operations for the scenarios we
consider.  Last, we show that \Name's controllers can perform upwards of 50
provisioning operations per second; given that provisioning occurs on
the order of tens of seconds, this is sufficient to support hundreds of tenants.

\section{Requirements and Related Work}
\label{s:challenges}

\begin{figure} 
\centering 
\includegraphics[width=0.9\columnwidth]{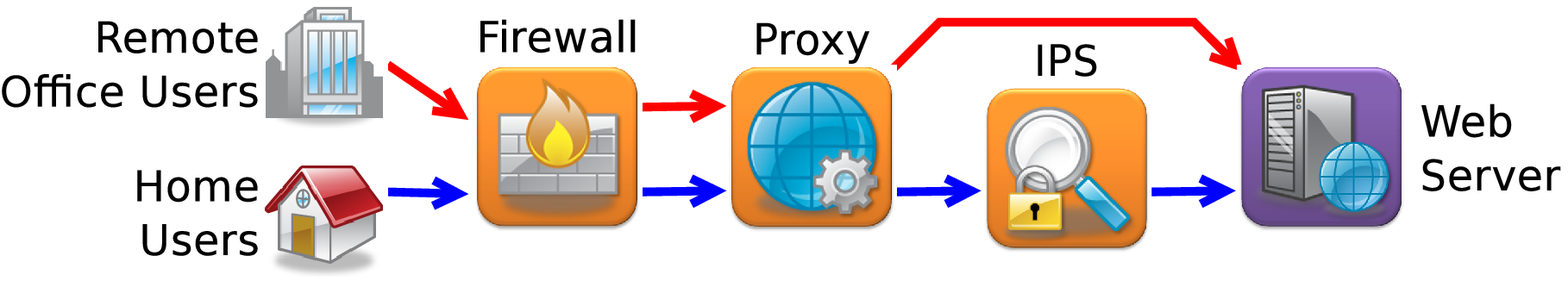} 
\compactcaption{Example \mb deployment with two chains (shown in red/blue).}
\label{f:example_chains} \end{figure}

Our goal is to build a \mb orchestration system that enables cloud tenants to (1)
{\em compose} rich, custom chains atop their \mb deployments. A chain is a sequence of \mbs that process a given traffic subset: e.g., an enterprise may require traffic
from a remote office to a company web server to pass through a firewall and
caching proxy (chain 1), and traffic from home users to pass through a firewall, proxy,
and intrusion prevention system (chain 2) (\figref{f:example_chains}); and (2)
automatically {\em provision} a suitable amount of resources for each chain to
optimally serve tenants' workloads. 

We argue that such a system must simultaneously meet the following requirements:

\begin{compactitem}
\item {\bf Correctness:} The physical realization of chains must correctly
  apply high level policies to traffic sub-streams.
\item {\bf Application-specific objectives and efficiency}:
 The system should 
  enable tenants to use the minimal amount of resources necessary to 
  realize application-specific service-level objectives (SLOs). 
\item {\bf Scalability:} The system should scale to
  hundreds-to-thousands of \mb chains from many tenants.
\end{compactitem}


Meeting these requirements is challenging on four key fronts.  ({\em i})  the
{\em closed nature} of third-party \mbs that makes it difficult to instrument
them; ({\em ii}) {\em diversity}, both in the nature of actions applied to
packet streams and in the amount of resources, such as CPU, memory, and network
bandwidth, consumed during packet processing;  ({\em iii}) the {\em shared
nature} of cloud environments; and ({\em iv}) the {\em dynamicity} arising from
tenant workload variations (and potential  \mb migration). 

\begin{figure}
\centering
\includegraphics[width=0.9\columnwidth]{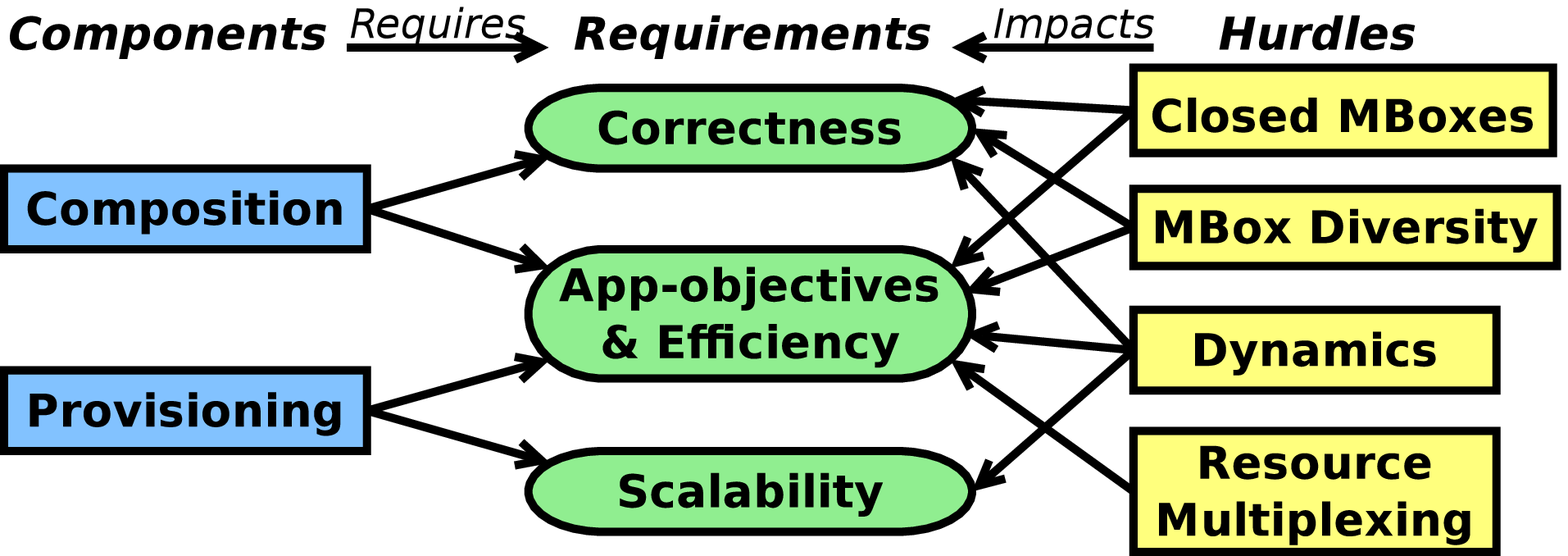}
\compactcaption{Interaction between the requirements for composition and provisioning and the challenges imposed by middleboxes and  cloud deployments}
\label{f:requirements_and_challenges}
\end{figure}

 Next, we explain how these factors make existing 
composition and provisioning solutions ineffective for meeting our requirements. In the
interest of brevity, we highlight only salient aspects and summarize the
interactions in \figref{f:requirements_and_challenges}.




\subsection{Composition}
\label{s:challenges_composition}


We first need a mechanism to enforce the appropriate steering of
traffic subsets across a chain of \mbs. 


A steering mechanism must meet two high-level correctness requirements.  First, at the granularity of an individual \mb, we note that many \mbs are {\em stateful} and require both directions of TCP sessions for correct functionality.  Thus, when a workload shift forces an {\em \mb deployment change} due to re-provisioning (\secref{s:challenges_provisioning}), forwarding rules must be updated to preserve {\em flow affinity}.
 Second, at the granularity of an individual \mb chain, we need to ensure that
a packet goes through the desired processing sequence.  However, many \mbs\
{\em mangle} packets by dynamically modifying packet fields; e.g., NATs rewrite
 headers and caching proxies terminate  connections.
Thus, the traffic steering rules must account for  modifications to
  determine the next hop \mb for mangled packets.



Existing approaches to steering fail to address at least one of these
 requirements. Some techniques (e.g., PLayer~\cite{joseph2008policyaware} and
SIMPLE~\cite{qazi2013simplefying}) are designed for {\em static} \mb deployments.
As such, they lack a nuanced approach for adjusting the fraction of traffic
assigned to specific \mb instances. They simply re-divide the flow space
among the available instances and replace existing steering rules; this may
cause some currently active flows to switch instances and violate the
affinity requirement\maybenixB{ (\tabref{t:composition_approaches})}.
Other techniques (e.g., consistent updates~\cite{reitblatt2011consistent,
wang2011openflow} or using per-flow
rules~\cite{casado2007ethane}\footnote{When a new flow starts, a central
controller installs flow-specific rules along the entire path for an \mb
chain.}) may be able to ensure that active flows maintain
affinity; however, they do not account for mangling.  



Tackling the  mangling problem is specially difficult because  \mbs are
 closed and proprietary.  Existing solutions fall
into two categories: ({\em i}) use flow correlations to reverse-engineer the
mangling relationships (e.g., SIMPLE~\cite{qazi2013simplefying}), or ({\em ii})
require \mbs to add persistent network handles to packets (e.g.,
FlowTags~\cite{fayazbakhsh2013flowtags} and service chaining
headers~\cite{boucadair2013differentiated, quinn2013network}). The former is
both expensive  (requiring multiple packets to be sent to the controller) and
error-prone (e.g., SIMPLE has 19\% error), while the latter needs \mb
modifications\maybenixB{ (\tabref{t:composition_approaches})}.

Efficiency implies minimal memory footprint in the switches and low latency
overhead in the controller that manages switch forwarding rules. Existing
solutions to this issue~\cite{wang2011openflow,yu2010scalable} apply
exclusively to simple routing and load balancing scenarios, and cannot
accommodate mangling \mbs and ensure affinity. As such, we need new schemes.

As we will see in \secref{s:composition}, \Name leverages the virtualized deployment to
engineer simpler more efficient solutions for both stateful forwarding and to
handle mangling \mbs.

\maybenixB{
\begin{table}
\centering
\small
\begin{tabular}{l|c|c|c|c}
    & {\bf Handles}     & {\bf Maintains}   & {\bf No \Mb}  & {\bf Minimal} \\
{\bf Framework} 
    & {\bf Mangling}    & {\bf Affinity}    & {\bf Changes} & {\bf Rules}   \\
\hline
PLayer~\cite{joseph2008policyaware} 
        & $\checkmark$  & X             & $\checkmark$  & ?             \\
SIMPLE~\cite{qazi2013simplefying} 
        & $\approx$     & X             & $\checkmark$  & $\checkmark$  \\
Consistent~\cite{reitblatt2011consistent, wang2011openflow} 
        & X             & $\checkmark$  & $\checkmark$  & $\checkmark$  \\
Per-Flow Rules
        & X             & $\checkmark$  & $\checkmark$  & X             \\
FlowTags~\cite{fayazbakhsh2013flowtags} 
        & $\checkmark$  & ?             & X             & $\checkmark$  \\
SC Header~\cite{boucadair2013differentiated, quinn2013network}
        & $\checkmark$  & ?             & X             & ?             \\
\end{tabular}
\compactcaption{Comparison of existing approaches for middlebox composition.}
\label{t:composition_approaches}
\end{table}}


\subsection{Provisioning}
\label{s:challenges_provisioning}



Two related  issues must be addressed to ensure that  tenant applications meet
their service-level objectives: ({\em i}) {\bf detection} to
determine where a resource bottleneck (or excess) exists in an \mb
chain, and ({\em ii}) {\bf provisioning} decisions on how/where resources need
 to be added (or removed) for the chain.  At a
high-level, existing approaches for detection can lead to inefficiency in the
\mb context, and  existing provisioning mechanisms can cause both inefficiency
and scalability issues.


\subsubsection{Resource bottleneck detection} 

A common approach (e.g., RightScale~\cite{rightscale}) is to monitor CPU and
memory consumption on individual virtual machines (VMs) and launch additional
VMs when some critical threshold is  crossed. Unfortunately, the shared nature
of clouds, and the unique and diverse resource profiles of \mbs together cause
this approach to both miss bottlenecks (impacting applications) and incorrectly
 infer bottlenecks (impacting efficiency). 
 For instance,  running multiple virtual \mbs on the
same host machine can lead to a memory cache bottleneck~\cite{dobrescu2012toward}.
Unfortunately, existing approaches will not lead to an appropriate scale-out
response in such cases, impacting application performance. 
 Similarly, \mbs that use polling to retrieve packets from the
NIC will appear to consume 100\% of the CPU regardless of the current traffic
volume~\cite{dobrescu2009routebricks}. In such cases, existing approaches will
cause spurious scale-out and reduce efficiency.  

Furthermore, these approaches do not consider network-level effects, which can
lead to further impact on applications and inefficiency. Indeed, 
\mb deployments in clouds are often impacted by transient network problems 
 that may cause application flows traversing \mbs to get backlogged and/or
timed-out~\cite{potharaju2013demystifying}. Persistent network
hotspots~\cite{benson2010network} can have a similar effect. 
 Thus, we may incorrectly conclude that the bottlenecks lie at \mbs
experiencing backlogs leading to ineffective scale-out response.

\subsubsection{Provisioning decisions} 

In a general setting,  we can use one of three options to alleviate
bottlenecks: (1) {\em horizontal scaling} to launch more instances based on
resource consumption estimates~\cite{dreger2008predicting}; (2) {\em migrate}  instances to less loaded
physical machines~\cite{wood2007blackbox}; and (3) choose appropriate {\em
placement} to avoid congested links (e.g.,~\cite{ballani2011oktopus,
lacurts2013choreo, meng2010improving, shrivastava2011applicationaware}).
Unfortunately, the dynamicity of clouds coupled with the varying
resource consumption of some \mbs renders existing techniques, and combinations thereof, inefficient and/or not scalable. 


 For instance, \mb resource consumption is quite diverse and workload
dependent~\cite{ghodsi2012multiresource}. Thus, scaling based on specific
resource indices may be inefficient and may not really improve end-to-end
application performance. Similarly, the bandwidth
consumption of \mbs can vary with the traffic mix (e.g., the volume of traffic
emitted by a WAN optimizer depends on how much the traffic is
compressed~\cite{anand2009redundancy}) and data center workloads can vary on
the order of tens of seconds to minutes~\cite{benson2010network}, so placement
decisions may only be optimal for a short period of time. 
Frequently invoking migration or scaling to accommodate these changes will
result in over-allocation of compute resources. It will also introduce
significant management overhead (for VM provisioning, virtual network set up,
        etc.) that can limit system scalability.

Thus, we need a systematic  framework for resource provisioning that ensures
\mb chain efficiency and system scalability in the face of \mb and cloud
dynamics.

\section{\Name Overview}
\label{s:central}

\begin{figure}
\centering
\includegraphics[width=0.8\columnwidth]{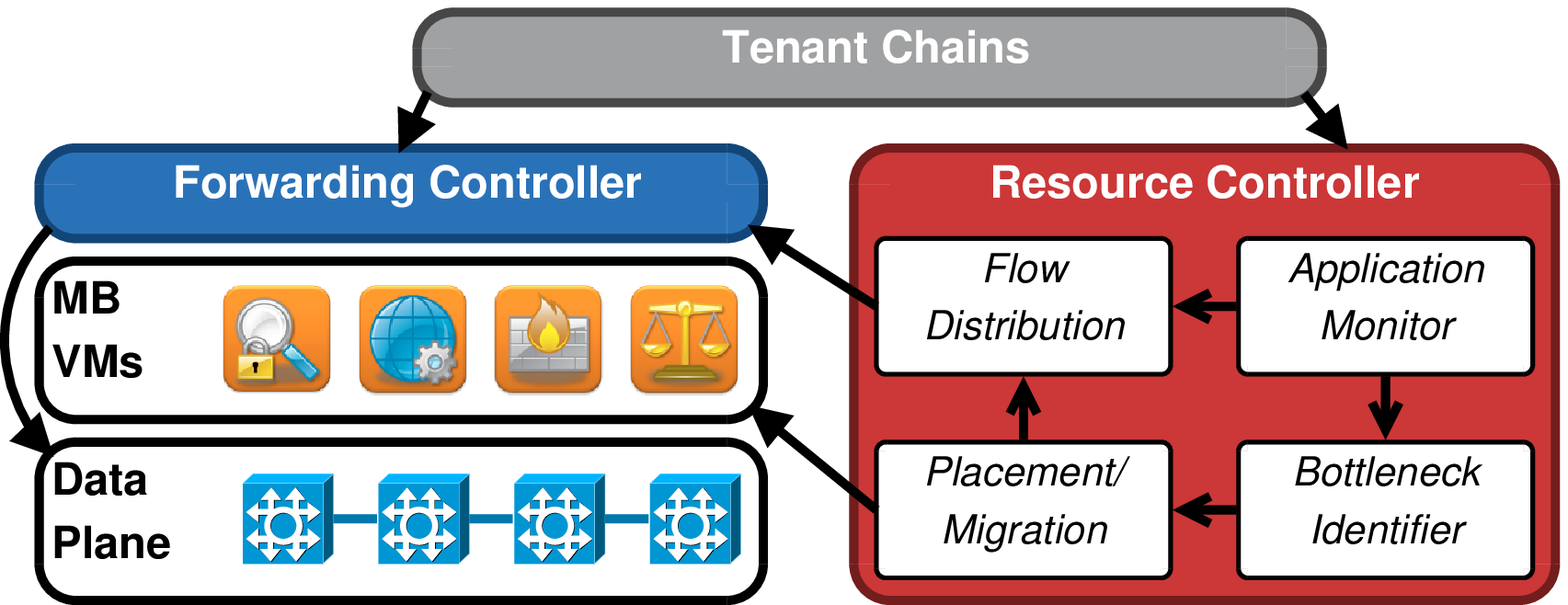}
\compactcaption{\Name overview}
\label{f:system_overview}
\end{figure}

\figref{f:system_overview} shows an overview of the \Name system with the interfaces 
 between the different components that we discuss briefly below. 
 \Name is a network-aware orchestration layer for virtual \mbs in clouds.  We
synthesize novel ideas with existing mechanisms to meet the provisioning and
composition requirements discussed in \secref{s:challenges}. 

\minisection{Tenant Input} Tenants provide a logical specification of their \mb
chains, including the applications/users associated with each chain and the VM
images to use for each \mb.

\minisection{Forwarding Controller} The data plane, composed of virtual
switches and tunnels, is programmed to forward traffic through \mb instances
according to chain specifications and computed flow distributions. 
 \Name  pre-processes  the  input chains to handle packet
mangling. It carefully manages individual flows uses a combination of tag-based
and per-flow rules to ensure correctness and efficiency (\secref{s:composition}).
 The forwarding controller also receives new provisioning  strategies output by the
 resource controller (see below), and  updates the data plane configuration
to ensure correct composition during and after these dynamic provisioning
decisions.



\minisection{Resource Controller} \Name uses end-to-end application performance
as a common, resource-agnostic indicator of an \mb chain facing performance
bottlenecks.  It monitors application performance and receives resource
statistics from the individual \mb and application VMs  as well as network
utilization statistics. Note that cloud providers already provide extensive
APIs to export this monitoring information to tenants~\cite{aws}. (As such, the design of
such a monitoring infrastructure is outside the scope of this paper.)

It uses a combination of three mechanisms---flow
distribution, horizontal scaling, and instance migration---applied at
progressively coarser time-scales to  identify and address bottlenecks
in an efficient and scalable
manner (\secref{s:provisioning}). 
Intuitively, flow distribution rebalances the  load at fine time-scales to \mb
replicas to address transient compute and network bottlenecks. This is
lightweight and can be applied often and in parallel across many chains, aiding
control plane scalability.  Horizontal scaling eliminates persistent compute
bottlenecks. Congestion-aware instance migration and horizontal scaling address
persistent network bottlenecks.  In both cases, scaled/migrated instances are
provisioned in a network-aware fashion, and flow distribution is applied to
improve efficiency.


In the next two sections, we present the detailed design of the \Name
forwarding and resource controllers.

\section{\Name Forwarding Plane}
\label{s:composition}


Our key insight to overcome the challenges discussed in \secref{s:challenges_composition} is that we can leverage unique features of the virtualized environment to engineer efficient composition approaches that handle mangling middleboxes and maintain affinity.
 

\subsection{Addressing \mb Mangling} 
\label{s:composition_mangling}
As discussed in \secref{s:challenges}, mangling/connection terminating (M/CT)
\mbs interfere with the ability to correctly forward packets to the
appropriate downstream \mbs. First, the identifiers required for selecting the
appropriate sequence of downstream \mbs may be obscured by an M/CT \mb's
packet modifications. Second, flow forwarding rules set up
in switches downstream from an M/CT \mb will cease to be valid when packet
headers change. 

\Name addresses the former issue by identifying potential sources of
forwarding ambiguity in the set of logical chains $\ChainArray$ provided by a
tenant and applying a correctness-preserving transformation to generate a
logically equivalent $\ChainArrayUnmang$ that is used for all subsequent
forwarding decisions. The latter is addressed by logically dividing a
chain $\chaininstance$ into subchains and installing per-flow forwarding
rules for a particular subchain when the first packet of a flow is emitted by
the first \mb in that subchain.

Prior to applying either mechanism, \Name must identify the set of \mbs in the
$\ChainArray$ that are potential M/CT \mbs using either: (1) operators'
domain expertise\footnote{While prior work~\cite{joseph2008policyaware}
required a detailed model of M/CT \mbs' mangling behavior, \Name only 
requires operators to identify which \mbs are M/CT \mbs.}, or (2)
monitoring the ingress/egress traffic of each \mb and checking if the output
packets fall in an expected region of the flow header
space~\cite{kazemian2012header}. 

\begin{figure}
\centering

\subfigure[Transformation result; identified subchains]{
    \includegraphics[width=0.75\columnwidth]{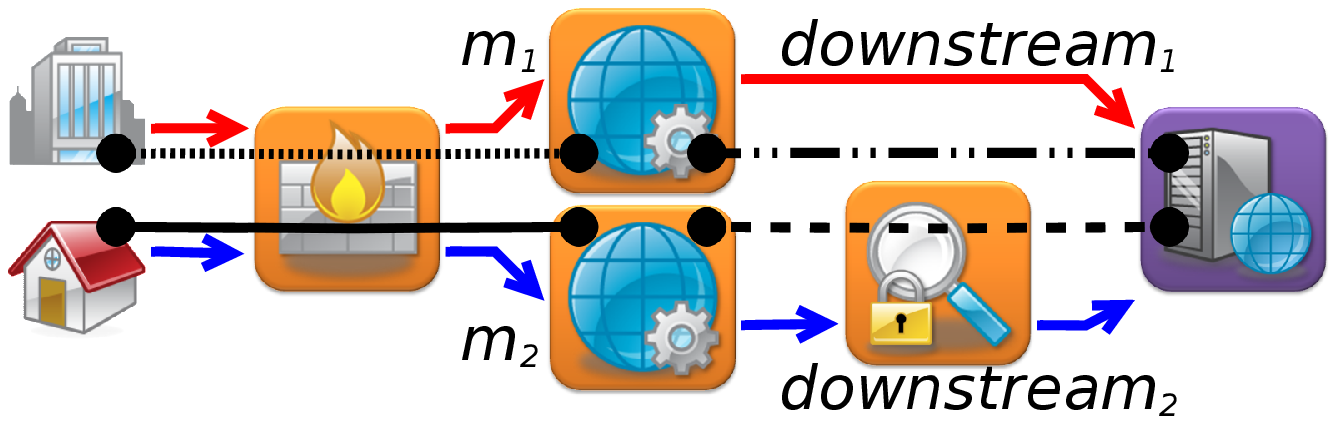}
    \label{f:example_unshare}
}
\subfigure[Paths and assigned tags for the blue chain] {
    \includegraphics[width=0.75\columnwidth]{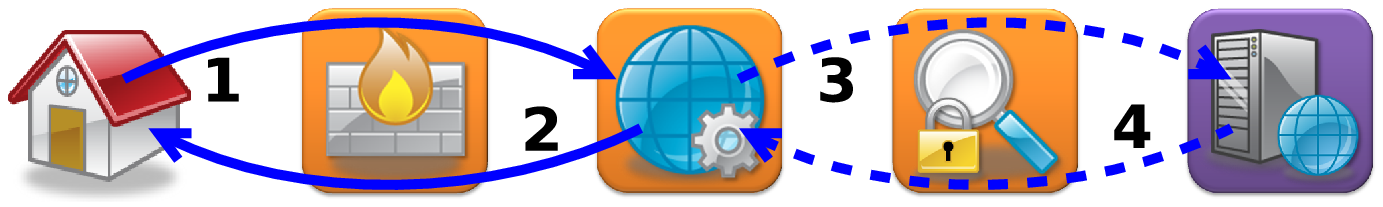}
    \label{f:example_paths}
}
\subfigure[Proactive and reactive tag-based rules]{
    \includegraphics[width=0.9\columnwidth]{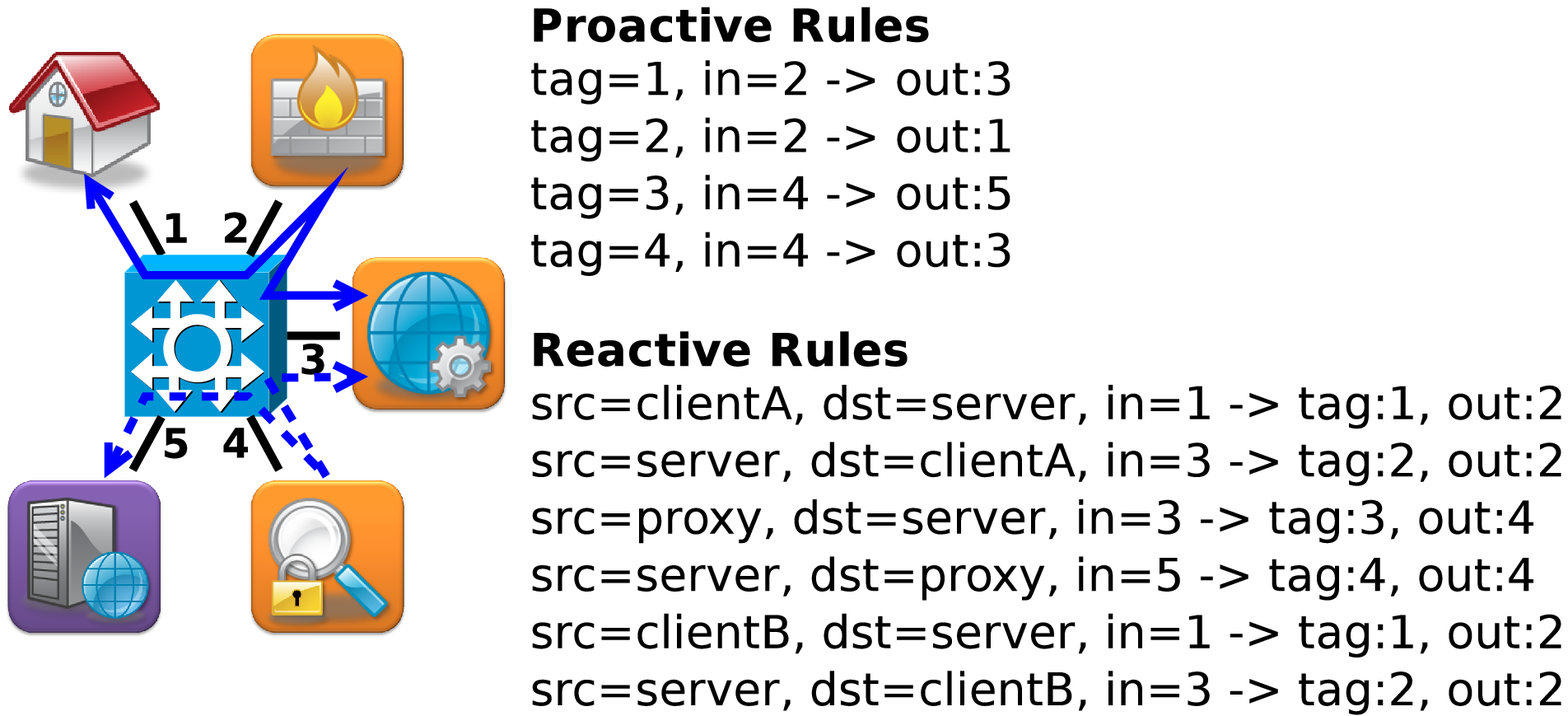}
    \label{f:example_rules}
}
\subfigure[Additional tag-based rules following provisioning]{
    \includegraphics[width=0.9\columnwidth]{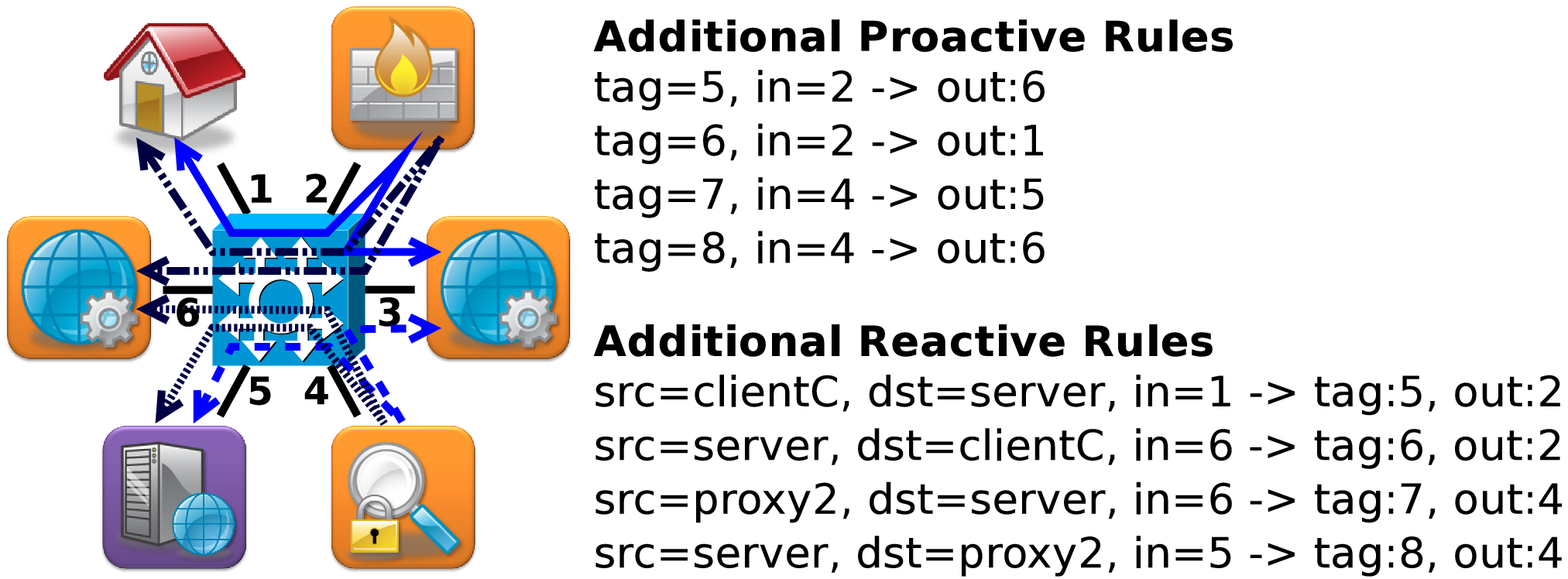}
    \label{f:example_rules_affinity}
}

\compactcaption{Steps for forwarding plane setup}
\label{f:example_composition}
\end{figure}

\minisection{Correctness-Preserving Transformation} Given the set of
chains $\ChainArray$, we can create a logical graph $G=\langle V,E
\rangle$, where each $v \in V$ is an \mb in a tenant specified logical
deployment and an edge $\middleboxinstance \rightarrow
\middleboxinstance'$ exists if there is a chain $\chaininstance$ with
the corresponding sequence. For each M/CT \mb $\middleboxinstance$, we
exhaustively enumerate all downstream paths in the graph $G$ starting
from $\middleboxinstance$; let this be
$\downpaths_{\middleboxinstance}$. Then, we create
$|\downpaths_{\middleboxinstance}|$ {\em clones} of
$\middleboxinstance$ with the clone $\middleboxinstance_i$ solely
responsible for the $i^\mathit{th}$ path in
$\downpaths_{\middleboxinstance}$. For example, in
\figref{f:example_unshare}, there are two downstream paths from the
proxy; thus we create two copies of the proxy. The intuition here is
that the appropriate path $i$ for a packet emitted by
$\middleboxinstance$ may be ambiguous due to packet changes made by
$\middleboxinstance$; by explicitly allocating isolated instances
$\middleboxinstance_i$ for each path $i$ we avoid any confusion due to
mangling. We rewrite each affected chain $\chaininstance \in
\ChainArray$ with the corresponding clone $\middleboxinstance_i$
instead of $\middleboxinstance$, and add the rewritten chain to
$\ChainArrayUnmang$.

We acknowledge that this transformation potentially increases the
number of \mb instances that need to be used: e.g., even if one proxy
instance was sufficient to handle the traffic load for both chains, we
would still need two clones---that are then provisioned
independently---to ensure correct composition. We believe that the
simplicity and correctness guarantees made without modifying \mbs, in
contrast to prior solutions~\cite{qazi2013simplefying,
  fayazbakhsh2013flowtags}, makes this tradeoff worthwhile. 

\label{s:composition_reactive}
\minisection{Setting up forwarding rules}
Given $\ChainArrayUnmang$ and knowledge of M/CT \mbs, we logically split each
chain into one or more {\em subchains}, with M/CT \mbs delineating the
subchains: e.g., the black lines in \figref{f:example_unshare} indicate
subchains. Conceptually, a subchain represents a logical segment where the
packet traverses the network without modifications to packet header fields
that uniquely identify a flow.

\Name needs to reactively set up flow rules when a packet for a new flow is
emitted by the first \mb (or client/server) in a subchain. The \Name
forwarding controller chooses one of the possible {\em instance paths} that
implement this specific subchain---i.e., an instance path contains a
particular instance of each \mb in the subchain. (The specific path will be
chosen using weighted round-robin with the weights determined by \Name' flow
distribution module described in \secref{s:distribution}.) The \Name
forwarding controller installs {\em exact match rules} in the virtual
switches to which the \mb (and client/server) instances in the path are
connected; the virtual switches themselves are connected by tunnels.
\Name also installs flow rules for the reverse flow in order to maintain the
flow affinity.  

\subsection{Maintaining Efficiency and Scalability}
The above  approach guarantees correctness in the face of mangling. However,
we can further optimize rule installation both in terms of the number
of rules required and the imposed load on the controller.
The main insight here is that we can proactively install forwarding rules in
virtual switches to forward traffic {\em within each subchain} using rules
that forward on the basis of {\em tags} included in packet headers.
Tags are set at the first instance of each subchain by reactive rules, as
described above.

Using tag-based rules (versus per-flow rules) for forwarding within subchains
reduces the total number of rules installed in virtual switches, leading to
faster matching and forwarding of packets~\cite{moshref2012vcrib}.
Additionally, proactively installing some rules reduces the number of rules
the \Name forwarding controller must install when new flows arrive, enabling
fast forwarding and controller scalability. 
\maybenix{We quantitatively show these performance benefits in
\secref{s:eval_forwarding}.}

Initially there is only one instance of each \mb in a subchain, and thus
only one possible instance path. We assign two tags to this path, one for the
forward direction and one for the reverse: e.g., \figref{f:example_paths}
shows the paths and tags for the two subchains associated with the blue
chain. A tag should uniquely identify both a subchain and a direction, so
each tenant has a single tagspace. 

\Name installs wildcard rules that match both the forward (or reverse) tag
and the virtual switch port of the prior \mb instance and output packets to
the virtual switch port for the next \mb instance on the forward (or reverse)
path: e.g., \figref{f:example_rules} shows the rules for the blue chain.  
Per-flow rules are reactively installed as described in
\secref{s:composition_mangling}, but only at the first element in each
sub-chain on the forward and reverse paths. 

In our prototype, we place tags in the type-of-service field in the IP
header, limiting us to 64 unique paths across all of a tenant's subchains;
recent IETF drafts suggest adding a special header in which such a
tag could be stored~\cite{boucadair2013differentiated,
quinn2013network}.\footnote{We assume the \mbs do not modify these tag header
bits, otherwise they would be treated as mangling \mbs.}

\subsection{Affinity in Face of Dynamics}

As additional \mb instances are provisioned, there become more possible paths
for a subchain. \Name allocates new forward and reverse tags just for the
new paths, and installs the corresponding rules. The tags and forwarding
rules for existing paths remain unchanged to ensure all packets of a flow
traverse the same set of \mb instances in both directions; this is important
for ensuring stateful \mbs operate correctly.
\figref{f:example_rules_affinity} shows the additional rules that would be
installed if the proxy was horizontally scaled; the rules shown in
\figref{f:example_rules} remain untouched.

\section{Provisioning}
\label{s:provisioning}

\Name' network-aware control plane closely manages chain performance
to satisfy application SLOs. 
 In order to balance the two requirements of {\em efficiency} (i.e., use minimal resources for each \mb chain while meeting application
objectives) and scalability (i.e., time/overhead of reconfiguration), we use the following multi-stage 
 approach (\figref{f:provisioning_process}):  

\begin{figure}[thbf]
\centering
\includegraphics[width=0.98\columnwidth]{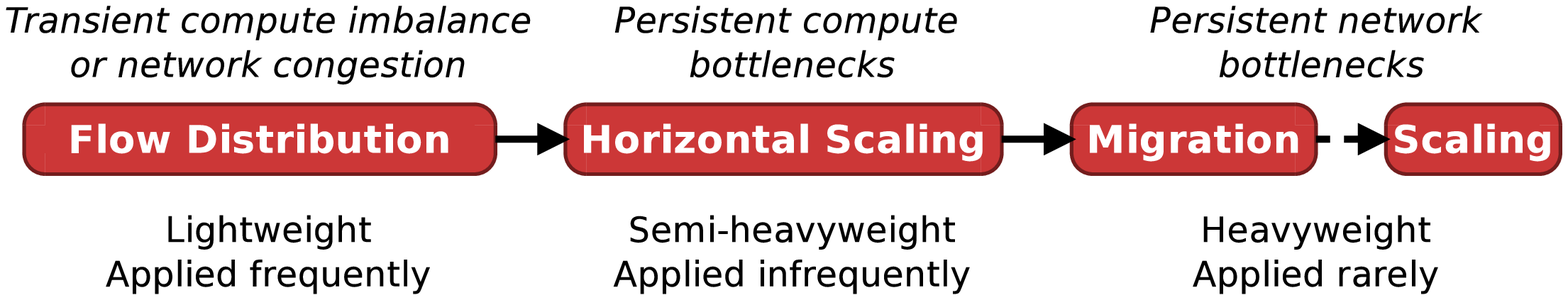}
\compactcaption{Process for detecting and addressing resource bottlenecks}
\label{f:provisioning_process}
\end{figure}

\begin{compactenumerate}

\item To determine the existence of a bottleneck, \Name leverages end-to-end
application-specific metrics, as suggested by some prior
approaches~\cite{bahl2007towards, wood2007blackbox}. Ultimately, these are
the true indicators of whether the chain is in need of extra resources; they can detect chain performance issues arising from fine-grained workload variations.


\item Given that the most common situation is for \mb chains to face
  transient compute or network bottlenecks, we design a light-weight
  flow distribution scheme - that can be invoked often without
  impacting scalability - to address such
  bottlenecks. The scheme does not need to know the nature
  of the bottleneck to address it -- whether compute or network, or even whether transient or persistent. It does not add new compute
  resources focusing instead on using them more efficiently
  (\secref{s:distribution}).

  When this scheme fails, we  identify if this  persistent is a compute or
network  bottleneck before taking appropriate measures.


\item We synthesize a suite of techniques to identify
  persistent compute bottlenecks. Each bottlenecked \mb is horizontally
  scaled by the minimal number of extra instances
  necessary (\secref{s:compute}). 

\item We use light-weight passive monitoring  to identify persistent
  network bottlenecks. Since addressing such bottlenecks can be
  tricky, we use a multi-pronged approach to ensure scalability. We first attempt to migrate instances in an inverse
  congestion-sorted order. If this does not help, we horizontally
  scale instances affected most by network congestion (\secref{s:network}).

\item When \#2 and \#3 fail, we fall back to horizontally scaling all
  \mbs by a fixed amount.

  In \#3, \#4, and \#5 above, we directly monitor application level
  metrics to identify if our decisions were effective, and we employ
  network-aware placement and re-invoke flow distribution so that the
  scaled/migrated instances are used in the most efficient fashion.

\end{compactenumerate}

Thus, our careful design systematically and accurately, addresses
performance bottlenecks, and ensures efficient and scalable operation.

\subsection{Flow Distribution}
\label{s:distribution}

Flow distribution adjusts the fraction of flows assigned to a specific
set of \mb instances so as to: ({\em i}) balance load on \mb instances
to accommodate variations in the rates, sizes, and types of flows
traversing the chain~\cite{dreger2008predicting} and ({\em ii})
control transient network congestion from changing load on network
links~\cite{benson2010network}.

We cast this problem as a linear program (LP).
Let $\chaininstance$ denote a specific chain and
$\numinchain$ denote the number of \mbs in that chain. The \mb (e.g.,
caching proxy, IPS) at position $\chainposition$ in the chain
$\chaininstance$ is denoted by $\chaininstance[\chainposition]$.
$\MiddleboxSet_{\chaininstance[\chainposition]}$ is the set of instances of
\mb $\chaininstance[\chainposition]$; we use
$\middleboxindex \in \MiddleboxSet_{\chaininstance[\chainposition]}$ to
specify that $\middleboxindex$ is an instance of this \mb.
$\Volume_\chaininstance$ denotes the total volume (bytes) of traffic that
must traverse chain $\chaininstance$\maybenix{; we discuss in
\secref{s:metrics} how we determine the value of
$\Volume_\chaininstance$}.

Our goal is to split the traffic across the instances of each \mb such
that: (a) the processing responsibilities are distributed roughly
equally across them, and (b) the aggregate network footprint is
minimized. In contrast with prior works that focus on equalizing \mb
load~\cite{sekar2010network, heorhiadi2012new, qazi2013simplefying},
our formulation has the benefit of eliminating (or reducing) compute {\em
  and} network bottlenecks, and reducing the likelihood of them
occurring in the future.


We need to determine how the traffic is routed
between different \mb instances. Let
$\Split(\chaininstance,\middleboxindex,\middleboxindex^{\nextto})$
denote the volume of traffic in chain $\chaininstance$ being routed
from \mb instance $\middleboxindex$ to \mb instance
$\middleboxindex^\nextto$. As a special case,
$\Split(\chaininstance,\middleboxindex)$ denotes traffic steered to
the first \mb in a chain from a source element.\footnote{For clarity,
  we focus only on the forward direction of the chain.}

$\Footprint(\middleboxindex \rightarrow
\middleboxindex')$ denotes the network-level cost between two
instances.  In the simplest case, this is a binary variable---1 if the two
\mb instances are on machines in different racks and 0 otherwise.
We capture current available bandwidth: Let $\rackinstance$ denote a specific
rack, and $\middleboxindex \in \rackinstance$ indicate that $\middleboxindex$
is located in that rack. The current bandwidth available between two racks
$\rackinstance$ and $\rackinstance'$ is denoted by $\Available(\rackinstance,
\rackinstance')$. \maybenix{\secref{s:metrics} describes how we measure
available bandwidth in a scalable fashion.}

\begin{figure}[t]
\begin{scriptsize}
  \fbox{
  \begin{minipage}[t]{0.95 \columnwidth}
Minimize 
\begin{align}
& \sum_{\chaininstance}
\sum_{\chainposition=1}^{\numinchain-1}
\sum_{ \substack{\middleboxindex,\middleboxindex^\nextto s.t \\
\middleboxindex \in \MiddleboxSet_{\chaininstance[\chainposition]}; \middleboxindex^\nextto \in \MiddleboxSet_{\chaininstance[\chainposition+1]} \\
 } }	\Footprint(\middleboxindex,\middleboxindex^\nextto)
\times
\Split(\chaininstance,\middleboxindex,\middleboxindex^\nextto)  \label{eq:footprint} \\
& \mathit{subject\ to} \nonumber \\
& \forall \middleboxindex,
 \forall \chaininstance,\ \mathit{s.t.}\ \middleboxindex \in \MiddleboxSet_{\chaininstance[\chainposition]}\ \&\ \chainposition
 > 1: \nonumber \\
 & \sum_{\middleboxindex^\prevto:
\middleboxindex^\prevto \in \MiddleboxSet_{\chaininstance[\chainposition-1]} }
\Split(\chaininstance,\middleboxindex^\prevto,\middleboxindex)  = \sum_{\middleboxindex^\nextto:
\middleboxindex^\nextto \in \MiddleboxSet_{\chaininstance[\chainposition+1]} }
\Split(\chaininstance,\middleboxindex,\middleboxindex^\nextto)  \times \Gain(\chaininstance,\chainposition)
\label{eq:conservation} \\
&	\forall \chaininstance: \sum_{\middleboxindex:
\middleboxindex \in \MiddleboxSet_{\chaininstance[1]}} \Split(\chaininstance,\middleboxindex)   =  \Volume_\chaininstance \label{eq:coverage}\\
& \forall r,r':
\sum_{c}
\sum_{\substack{i,i' s.t. \\
i \in r; i' \in r' \\
}} f(c,i,i') \le b(r, r') \label{eq:bw} \\
&	 \forall \middleboxindex:
\sum_{ \substack{\chaininstance: \middleboxindex \in \MiddleboxSet_{\chaininstance[\chainposition]}; \chainposition \ne 1}
 } \sum_{ \substack{\middleboxindex^\prevto:
\middleboxindex^\prevto \in \MiddleboxSet_{\chaininstance[\chainposition-1]}}}
 \Split(\chaininstance,\middleboxindex^\prevto,\middleboxindex)  \nonumber \\
& + \sum_{\chaininstance: \middleboxindex \in  \MiddleboxSet_{\chaininstance[1]}
 } \Split(\chaininstance,\middleboxindex)   \approx \sum_{\chaininstance:
\middleboxindex \in \chaininstance; \middleboxindex \in \MiddleboxSet_\chaininstance[\chainposition]}
 \frac{\Volume_\chaininstance}{|\MiddleboxSet_\chaininstance[\chainposition]|}  \times \Pi_{l=1}^{\chainposition} \Gain(\chaininstance,l)
\label{eq:loadbalance}
  \end{align}
\end{minipage}
}
\compactcaption{LP formulation for the flow distribution problem.
The $\approx$ term in the last equation simply represents that we have some leeway
 in allowing the load to be within 10--20\% of the mean.}
\label{fig:lp-split}
\end{scriptsize}
\end{figure}

\minisection{LP Formulation}
Figure~\ref{fig:lp-split} formalizes the flow distribution
problem that \Name solves. Eq~\eqref{eq:footprint} captures the
network-wide footprint of routing traffic between instances of the
$\chainposition^\mathit{th}$ \mb in a chain to the
$\chainposition+1^\mathit{th}$ \mb in that chain.  For completeness, we
consider all possible combinations of routing traffic from one instance to
another. In practice, the optimization will prefer combinations that
have low footprints.

Eq~\eqref{eq:conservation} models a byte conservation principle. For
each chain and for each \mb in the chain, the volume of traffic entering
the \mb has to be equal to the volume exiting it. However, since \mbs may
change the aggregate volume (e.g., a WAN optimizer may compress traffic), we
consider a generalized notion of conservation that takes into account a
gain/drop factor $\Gain(\chaininstance,\chainposition)$: i.e.,
the ratio of ingress-to-egress traffic at the position $\chainposition$
for the chain $\chaininstance$.
\maybenix{\Name computes these ratios based on virtual switch port statistics
(\secref{s:metrics}).}

We also need to ensure that each chain's aggregate traffic will be processed;
we also model this coverage constraint in Eq~\eqref{eq:coverage}. We also need
to ensure that total chain traffic across any two racks does not exceed the available bandwidth
between the two racks; we model this bandwidth constraint in Eq~\eqref{eq:bw}.
Finally, we use a general notion of load balancing where we can allow for some
leeway; say within 10-20\% of the targeted average load (Eq~\eqref{eq:loadbalance}).

\subsection{Identifying and Addressing Bottlenecks}
\label{s:bottleneck}

There are cases when flow distribution will be insufficient to improve end-to-end performance: e.g., when all instances of an \mb, or all paths to those instances, are heavily loaded, or when network/\mb loads are such than a redistribution is simply infeasible. In such cases, \Name is forced to identify the type of bottleneck (compute or network) that exists and address it. To overcome the challenges outlined in \secref{s:challenges}, we adopt decouple the actions for dealing with the two types of bottlenecks: we focus on addressing compute bottlenecks first, followed by network bottlenecks.
 

\subsubsection{Compute Bottlenecks} 
\label{s:compute}

\Name leverages a combination of host-level and per-packet metrics 
 to determine whether a compute bottleneck exists, and for which \mbs. 
 Host-level metrics are used by 
 existing  scaling frameworks because these can be easily gathered from VMs~\cite{rightscale, wood2007blackbox}.  
 In addition to the host-level metrics, we rely on  
 the  packet processing time 
 as it can  capture {\em any} compute-related bottleneck,
including CPU, memory space/bandwidth, cache
contention~\cite{dobrescu2012toward}, and disk or network I/O.\footnote{
  This may not apply to  \mbs that do
not follow a ``one packet in, one packet out'' convention (e.g., a WAN
 optimizer), and in these cases we can only use traditional CPU and memory
utilization metrics.} 

 \Name declares an \mb instance to be bottlenecked if either: ({\em
i}) average per-packet processing time increased by at least a factor $\delta$
over a time window, or ({\em ii}) CPU or memory utilization exceeds a threshold
$\alpha$ and has increased by at least a factor $\beta$ over a sliding time
window.\footnote{The increase factor avoids constant scaling of \mbs which use
polling.} We select these thresholds heuristically based on 
 observing middlebox behaviors in controlled settings and varying 
 the offered load. \maybenix{\secref{s:metrics} discusses a scalable approach for
gathering these metrics.}



\label{s:scaling}

\minisection{Horizontal Scaling}
When compute bottlenecks are identified, \Name horizontally scales  each bottlenecked \mb and 
 adds more such instances of. (Our current implementation increases 
 only one instance at a time to avoid overprovisioning, but we could consider 
 batched increments as well.) Crucially, these instances must
be launched on machines that have, and likely will continue to have,
high available bandwidth to instances of other \mbs in the chain.  This helps
maximize the extent to which the new resources are utilized and minimizes
the need to perform migration (which can hurt scalability) or further scaling
(which can hurt efficiency) in the future.

We use a network-aware placement heuristic similar to
CloudNaaS~\cite{benson2011cloudnaas}: We try to place a new \mb instance in
the same rack as instances of the neighboring \mbs (or clients/servers) in
the chain; if the racks are full, we try racks that are two hops away,
before considering any rack. For each candidate rack, we calculate
the flow distribution (\secref{s:distribution}) as if the new instance was
placed there, and we choose the rack that results in the best objective value
(i.e., minimizes the volume of inter-rack traffic).

\maybenix{A bottleneck at one \mb in a chain may mask bottlenecks at other \mbs in
the chain: e.g., a bottleneck at the proxy in \figref{f:example_chains} will
limit the load on the IPS; when the proxy bottleneck is resolved, load on the
IPS will increase and it may become bottlenecked. Thus, we look
for compute bottlenecks multiple times and perform scaling until no \mbs in a
chain exhibit the conditions discussed above.}


\subsubsection{Network Bottlenecks}
\label{s:network}

Network bottlenecks may arise at any of the physical links that form the
underlay for the virtual links (VLs) connecting neighboring \mb instances
(i.e., the link from $\middleboxindex \in
\MiddleboxSet_{\chaininstance[\chainposition]}$ to $\middleboxindex' \in
\MiddleboxSet_{\chaininstance[\chainposition \pm 1]}$). Using active probing
to measure VLs' available bandwidth is not scalable: probing must occur
(semi-)serially to avoid measurement interference caused by multiple VLs 
sharing the same underlying physical link(s). Hence, \Name detects
bottlenecked VLs by passively monitoring the individual physical links that
underly VLs. This requires gathering metrics from all physical network switches and
identifying the physical links (e.g., using traceroute\footnote{Multipath
routing (e.g., ECMP) based on layer 4 headers may interfere with our ability
to do so; we leave this as an issue for future work.}) that form each
VL\maybenix{;
these tasks can easily be parallelized, as described in \secref{s:metrics}}.
A VL is bottlenecked when the minimum available bandwidth across the
physical links that form the VL is less than a threshold $\delta$.

\label{s:moving}

\minisection{Instance Migration}
When a network bottleneck is identified, \Name migrates affected \mb
instances to less congested network locations. Since migrations are costly
operations---involving, in our prototype, the instantiation of a new \mb
instance with a more optimal placement and the termination of the instance
affected by the network bottleneck(s)---performing the minimum number of
migrations is crucial to maintaining system scalability. For this reason,
\Name first migrates \Mb instances with the highest number of incident
congested VLs and measures the migration's impact on end-to-end application 
performance before performing additional migrations. An \mb instance's new
location is selected using the placement heuristic described in
\secref{s:scaling}, with the added requirement that the available bandwidth
between the \mb instance being migrated ($\middleboxindex \in
\MiddleboxSet_{\chaininstance[\chainposition]}$) and the instances
of neighboring \mbs in the chain ($\middleboxindex' \in
\MiddleboxSet_{\chaininstance[\chainposition \pm 1]}$) must be
greater than the current bandwidth consumed by $\middleboxindex$ times some 
factor $\rho$.

In some cases, the network bottleneck(s) impacting an \mb instance may arise
predominantly due to heavy traffic involving the instance itself (e.g.,
bandwidth needs may outstrip compute needs~\cite{ghodsi2012multiresource}).
In these cases, and cases where all portions of the cloud network are
congested, the network-aware placement routine will not yield a feasible
solution. We address this situation by horizontally scaling the instance
using the technique described in \secref{s:scaling}. This causes chain
traffic to be spread among more \mb instances, reducing the network bandwidth
needed by an individual instance and eliminating any network bottlenecks. It
also causes underutilization of compute resources on the affected
instance(s), which reduces efficiency.

\subsection{De-provisioning}
\label{s:deallocating}
To maintain efficiency, \Name also eliminates excess compute resources.
Excesses are identified by looking for \mb instances whose average per-packet
processing time has dropped by at least a factor $\delta'$ over a time
window. To avoid sudden violations of application SLOs, supposedly unneeded
\mb instances are removed from service (but not yet destroyed) one at a time;
flow distribution (\secref{s:distribution}) is invoked to rebalance load
among the remaining instances. If the SLOs of {\em all} applications
associated with the \mb chain are satisfied, then the instance is marked for
permanent removal; otherwise, the instance is immediately restored to service
(using the old flow distribution values). An instance is fully destroyed only
after all flows traversing it have finished (or timed-out).

\maybenix{\subsection{Data Collection} 
\label{s:metrics}

The \Name resource  controller relies on many metrics when making provisioning
decisions, but all of the metrics can be gathered in a scalable fashion by
leveraging distributed monitoring agents and a centralized object
store (e.g., ~\cite{lakshman2010cassandra}). Most applications already log
end-to-end performance measures for other purposes; this can simply also be
reported to \Name. The volume of traffic traversing an \mb chain and the
gain/drop factor and average per-packet processing time of \mbs in the chain
can be captured by querying the port statistics from each virtual switch. An
agent running on each machine can perform such queries, as well as report
VMs' CPU and memory utilization. Lastly, a collection of sensors can poll
port statistics from physical switches using SNMP.
}

\section{Implementation}
\label{s:implement}

\begin{figure}
\centering
\includegraphics[width=0.9\columnwidth]{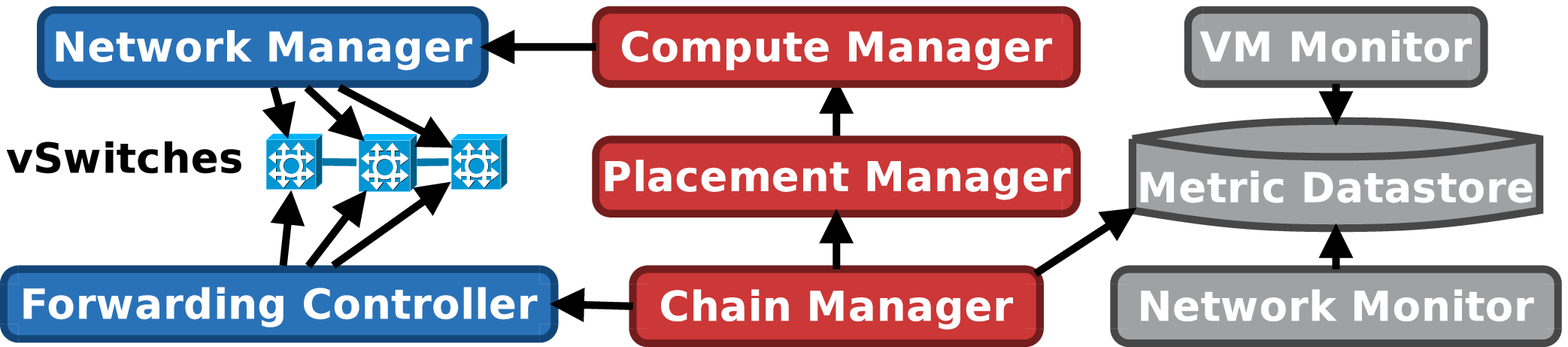}
\compactcaption{\Name prototype implementation}
\label{f:implementation}
\end{figure}

We have implemented a full featured \Name prototype consisting of several
components (\figref{f:implementation}). \Name' modular design makes
it easy to scale individual components as the number of tenants and the
size of the cloud increases.

\minisection{Forwarding Controller and Data Plane}
The \Name data plane is a configurable overlay network realized through
packet encapsulation and SDN-enabled software switches. Each machine runs an
Open vSwitch~\cite{openvswitch} bridge to which the virtual NICs for the VMs
running on the machine are connected. The {\em network manager} establishes a
full mesh of VXLAN tunnels for each tenant, connecting the vSwitches to which
a tenant's VMs are connected. 

The forwarding controller is implemented as a module ($\approx$2400 lines of
Java code) running atop the Floodlight OpenFlow
Controller~\cite{floodlight}. Floodlight handles communication with the
vSwitches, and the forwarding module interfaces with the resource controller
and network manager using Java Remote Method Invocation (RMI).

\minisection{Resource Controller}
The {\em chain manager} ($\approx$6000 lines of Java) forms the core of the
resource controller. It monitors the performance of tenant applications
(through queries to the {\em metric datastore}) and executes the provisioning
process (\figref{f:provisioning_process}) when SLO thresholds are crossed.
Flow distribution is computed using CPLEX~\cite{cplex} and applied by the
{\em forwarding controller}; scaling and migration decisions are applied by
the {\em placement manager} ($\approx$ 3300 lines of Java). When placement
decisions are made, the {\em compute manager} communicates with the
Xen~\cite{xen} hypervisor to launch VMs.

The metrics required for provisioning decisions reside in the {\em metric
datastore}, currently a simple in-memory data store written in Java.  A {\em
VM monitor} runs on each machine and reports the CPU, memory, and network
metrics for running VMs based on output from Xen. The {\em network
monitor} queries port statistics from physical switches using SNMP and
reports current utilization for each physical link. 

\section{Evaluation}
\label{s:eval}

We use a combination of testbed (\secref{s:eval_testbed}) and simulation experiments to evaluate
\Name' ability to satisfy our key requirements (\secref{s:challenges}):
 \begin{compactitem}
 
\item First, we examine the ability of the \Name forwarding plane to correctly and
efficiently realize complex \mb chains in the presence of dynamics.  (\secref{s:eval_forwarding}.)

\item Second,
we measure how adequately and efficiently \Name' provisioning mechanisms
satisfy application objectives, highlighting how key aspects of our
provisioning process contribute to the observed outcomes. (\secref{s:eval_provisioning}.)

\item Last, we establish
the scalability of \Name' forwarding and resource controllers. (\secref{s:eval_scalability}.)

\end{compactitem}

\subsection{Testbed Setup}
\label{s:eval_testbed}

The majority of our evaluation is conducted in a small cloud testbed.
The testbed consists of 36 machines (quad-core 2.4GHz, 2.67GHz, or 2.8GHz
CPU and 8GB RAM) deployed uniformly across 12 racks. Each machine, running
Xen and Open vSwitch, has 3 VM slots.  The racks are connected in a simple
tree topology with 12 top-of-rack (ToR) switches, 3 aggregation switches,
and 1 core switch.

We run a variety of \mbs, including an IPS (Suricata~\cite{suricata}),
redundancy eliminator (SmartRE~\cite{anand2009smartre}), and two synthetic
Click-based~\cite{kohler2000click} \mbs: {\em passthrough} forwards packets
unmodified, and {\em mangler} rewrites packets' source IP (in the forward
direction) and destination IP (in the reverse). We also run Apache web server
and a custom workload generator. The workload generator runs 8 client threads
that draw tokens from a bucket filled at a specified rate. For each token, a
client thread issues an HTTP POST of a fixed size (0.2KB, 10KB, 50KB, or
100KB), and receives a reply of the same size, with a request timeout of 1
second. Client threads block if no tokens are available; the number of
outstanding tokens (maximum 100) indicates unmet demand.

We generate background traffic between pairs of machines using iperf to 
send UDP packets at a fixed rate.

\begin{figure}
\centering
\includegraphics[width=0.9\columnwidth]{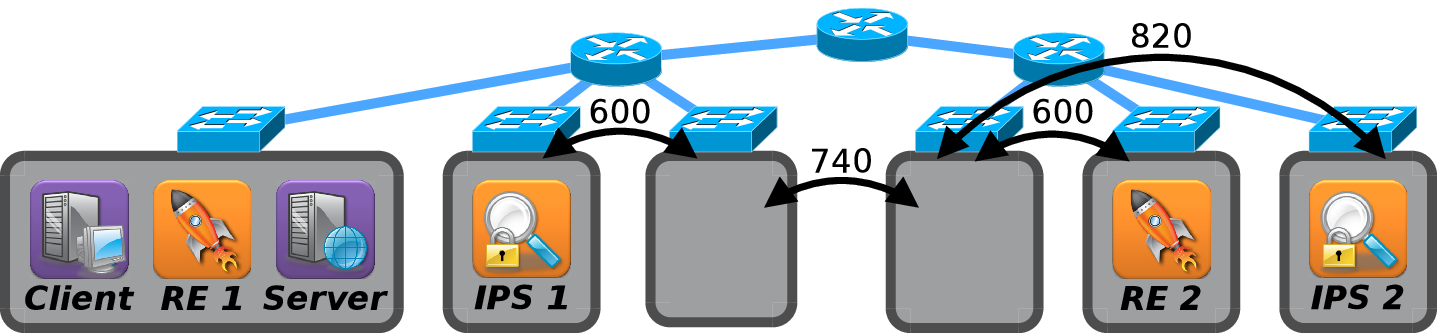}
\compactcaption{Initial instance placement and background traffic patterns
    for provisioning evaluation}
\label{f:provisioning_eval_setup}
\end{figure}

\begin{figure}
\centering
\includegraphics[width=0.9\columnwidth]{provision_stratos_events.epsi}
\compactcaption{Timeline of load changes and provisioning actions}
\label{f:provision_stratos_events}
\end{figure}

\subsection{Composition Efficiency and Correctness}
\label{s:eval_forwarding}
We first examine the ability of \Name' forwarding controller to efficiently 
 realize complex \mb chains in the presence of dynamics.  

\minisection{Efficiency} First, we measure the inflation in per-flow request
completion time per-flow caused by the need to install per-flow rules. We
construct a chain consisting only of a workload generator and web server
(placed on different physical machines), and forward traffic between them  %
with and without the \Name' forwarding plane.  We generate 100 requests/sec
for 10 minutes and vary the request sizes for each run between 0.2KB and 100KB.
Comparing the average request completion time with and without \Name'
forwarding plane, we observe that the average inflation is no more than 1ms per flow.
 (In general,  for a  chain with  $N$ mangling \mbs, the increase 
 in latency will be  $N$~ms  higher in the worst case.)

We also benchmarked the performance of our tag-based forwarding and
found that the overhead is minimal; this is consistent with prior reports on
the performance of the OpenvSwitch dataplane~\cite{ovs}.

\minisection{Correctness} To evaluate affinity, we pick a chain with a
workload generator, passthrough \mb, mangler \mb, and web server. Every two
minutes, we add an instance of one of the \mbs and compute a new flow
distribution. We cycle through the \mbs in round robin order. We repeat the
process for two rounds. Throughout an experiment, we generate a constant
workload of 50KB requests at a rate of 50 requests/sec.

We run tcpdump on each \mb during an experiment and afterwards feed each
trace to Bro~\cite{paxson1998bro} which produces a log of all connections.
For each flow (which Bro identifies with a hash based on key packet header
fields), we compare the flags, byte counts, and packet counts across the
connection logs for the workload generator, web server, and each \mb. Since
the mangling \mb modifies packet headers, and hence results in different
connection hashes, we only compare the connection logs from each half of the
chain. If affinity were to be broken, discrepancies can arise in the
logs; e.g., connections whose packets are split between multiple \mb
instances will appear in the logs of multiple \mbs and the log entries will
indicate that not all key packets (e.g., SYN and FIN) were seen by a given
instance.

We conducted the above experiment on two chains---workload generator,
passthrough, mangler, server and workload generator, passthrough,
passthrough, server---and found no such discrepancies in the logs, thus
affirming that \Name' steering ensures affinity.

 We note that our approach of unsharing \mbs to handle mangling is correct by
construction; nevertheless, we also validated the correctness 
 using a similar logging approach. We do not discuss this in the 
 interest of brevity. 



\subsection{Provisioning Adequacy and Efficiency}
\label{s:eval_provisioning}

\begin{figure*}
\centering
\subfigure[Stratos]{
    \includegraphics[width=0.72\columnwidth]{provision_stratos_throughput.epsi}
    \label{f:provision_stratos_throughput}
}
\hspace{-1em}
\subfigure[\NoFlow]{
    \includegraphics[width=0.7\columnwidth]{provision_noflow_throughput.epsi}
    \label{f:provision_noflow_throughput}
}
\hspace{-1em}
\subfigure[\MonitorVMs]{
    \includegraphics[width=0.7\columnwidth]{provision_monitorvms_throughput.epsi}
    \label{f:provision_monitorvms_throughput}
}
\hspace{-1em}
\subfigure[\UniformFlow]{
    \includegraphics[width=0.7\columnwidth]{provision_uniformflow_throughput.epsi}
    \label{f:provision_uniformflow_throughput}
}
\compactcaption{Provisioning decisions and application throughput and backlog
   under various systems}
\label{f:provision_throughput}
\end{figure*}


We use a single chain consisting of a workload generator, an RE \mb, an IPS \mb, 
and a web server. Two initial instances of each \mb are placed at fixed
locations, as shown in \figref{f:provisioning_eval_setup}, to simulate
placements that might occur in a cloud with existing tenants. The workload
we use (the {\em Demand} line in \figref{f:provision_stratos_throughput})
starts with 90 requests/sec and increases by 10 requests/sec at varying
frequencies to reach an ending rate of 175 requests/sec after 9
minutes; the size of each request is 100KB.

Throughout the experiment, we apply three different background traffic
patterns, shown in \figref{f:provisioning_eval_setup}: ({\em i}) at
experiment start, 600Mbps of traffic is exchanged between a pair of racks
under each of two aggregation switches; ({\em ii}) 75 seconds into the
experiment, the background traffic switches to 740Mbps between a rack
under each of the two aggregation switches, and this lasts for 1 minute; 
({\em iii}) roughly 6.75 minutes into the experiment, the background traffic
switches to 820Mbps between a pair of racks under one of the aggregation
switches.

\minisection{An illustrative run} The provisioning actions taken by
\Name, along with workload and link load changes are shown in
\figref{f:provision_stratos_events}. \Name leverages each of its
provisioning mechanisms at appropriate points in the scenario: flow
distribution occurs when transient network load shifts (and before all
other provisioning action), scaling occurs following a significant
increase in demand, and migration occurs when a persistent network
load is introduced.  We validate each of the actions below in our
discussion of how well \Name addresses application objectives and
efficiency. 

To understand how different pieces contribute to \Name' performance, we
compare \Name against three alternative designs that progressively exclude
specific  \Name components: (1) {\em \NoFlow}, a system that does not
proactively use flow distribution to alleviate bottlenecks but only invokes it
after scaling or migration events; (2) {\em \MonitorVMs}, which is similar to
\NoFlow, except that it only monitors the CPU, memory, and network link
consumption at VMs to identify bottlenecks, and initiates scaling (it does not
consider network effects beyond access links of VMs, and hence it does not try
migration); and (3) {\em \UniformFlow}, which is similar to \MonitorVMs, except
flows are uniformly distributed across \mb instances (as opposed to invoking
\Name' flow distribution) following horizontal scaling.

\minisection{Application Objectives} 
 While the design of \Name is quite general and can accommodate
 many different types of application SLOs (i.e., 
 in our
evaluation, we consider two such metrics -- throughput and request
backlog -- shown in \figref{f:provision_stratos_throughput} to
illustrate \Name' ability to satisfy application objectives.
 Specifically, we configure \Name to initiate the provisioning process
(\figref{f:provisioning_process}).
whenever average request latency exceeds 45ms
and/or the backlog exceeds 10 requests for at least a 10 second time period.


  \figref{f:provision_stratos_throughput} shows that \Name'
  provisioning actions restore application performance to acceptable
  levels relatively quickly: within 10 secs during the transient link
  load change (that occurs at 75 secs) and the demand spike (that
  occurs at 155 secs), and within $\approx$ 50 secs when the
  persistent link load change occurs (at 405 secs). The response takes
  longer in the latter case because migration is more heavyweight than
  flow distribution or scaling; it takes 45 secs to perform a VM
  launch (35 secs) and termination (10 secs), plus there is a 20 sec
  delay between flow distribution and detection of network bottlenecks
  while \Name waits to see if flow distribution was sufficient to
  address the bottleneck. Overall, with \Name, application requests
  served closely tracks application demand.



Excluding \Name components leads to inefficiency and/or inability to
meet application objectives. Consider the time period with transient
network load: \NoFlow and \MonitorVMs have clear gaps between 
 the demand and served load, and a full backlog. 
 (As noted in \secref{s:eval_testbed}, the maximum backlog is 100
requests.) \MonitorVMs has no
appropriate mechanism to address this bottleneck at all, and
application performance suffers. Without flow distribution as a
first-order provisioning option, \NoFlow attempts horizontal scaling.
While this succeeds eventually, request have backlogged in the
interim, \NoFlow ends up using more instances than absolutely
necessary, and management overhead (to launch a VM) is higher.

There is no bottleneck with \UniformFlow at this point in the scenario
because the starting flow distribution sends only half of the
application workload over the link with the transient load, and there
is sufficient capacity remaining on the link to handle this fraction
of the load. For comparison, \Name' initial flow distribution sends
the entire workload across the affected link, but \Name invokes flow
distribution again when the transient network load occurs to reduce
this load to half, resulting in the same situation as \UniformFlow.

Now consider the time period with persistent congestion. Recall that
such bottlenecks occur infrequently~\cite{benson2010network}. \Name'
backlog is better than all alternatives except \NoFlow. In contrast
with \Name, which first attempts flow distribution for scalability
reasons, \NoFlow addresses the network bottleneck directly by invoking
migration. View across the timeline, \Name is better in all respects
than \NoFlow: it invokes strictly fewer heavy weight operations (such
as VM launch), it results in lesser backlog on average, and it uses
resources at least as efficiently (a topic we will cover in more
detail shortly).




We found that \Name' performance is similarly competitive with
respect to application latency as well. We omit the results for
brevity.


\minisection{Efficiency} We examine \Name' provisioning efficiency
from three perspectives: number of \mb instances required, compute
resource utilization, and network link utilization.

\begin{table}
\centering
\small
\begin{tabular}{l|c|c|c|c}
& {\bf \# \mb} & {\bf Avg IPS CPU} & 
    \multicolumn{2}{c}{\bf Avg Link Utilization} \\
{\bf System} & {\bf Instances} & {\bf Utilization} & 
    {\bf ToR-Agg} & {\bf Agg-Core} \\
\hline
\Name & 5 & 70\% & 128Mbps & 144Mbps \\
\NoFlow & 6 & 65\% & 128Mbps & 165Mbps \\
\MonitorVMs & 7 & 53\% & 145Mbps & 146Mbps  \\
\UniformFlow & 6 & 64\% & 145Mbps & 188Mbps  \\
\end{tabular}
\compactcaption{Efficiency under various systems}
\label{t:provision_efficiency}
\end{table}

The number of \mb instances used to satisfy application objectives has a
direct impact on the costs (for cloud infrastructure, \mb licensing, etc.)
incurred by tenants. \Name is highly effective at optimizing this
metric. The RE \mb is never scaled or migrated in any of our
experiments, as the packet processing capacity of a single RE instance is
$\approx$3x higher than a single IPS instance. In contrast, the IPS \mb is
horizontally scaled once with \Name and 2-3 times with the other system
designs. The extra \mb instances launched by the other systems
(\tabref{t:provision_efficiency}) hurt efficiency.

Directly related to instance count is the utilization of each
instance, where higher is better as it indicates that \mb resources
are being used effectively. \tabref{t:provision_efficiency} shows the
average CPU utilization of the IPS instances. With \Name the average
utilization (70\%) is 15\% less than our CPU utilization threshold for
compute bottlenecks (85\%). In contrast, other systems have 5\% to
17\% lower average CPU utilization compared to \Name.

Finally, network link utilization indicates the likelihood of future network
load changes inducing bottlenecks that affect the \mb chains.
\tabref{t:provision_efficiency} shows the average bandwidth the chain
utilizes on top-of-rack switch to aggregation switch links (4 links are
used) and aggregation switch to core switch links (2 links are used). We
observe that the both the ToR-Agg and Agg-Core links have the lowest utilization with 
 \Name,  while \UniformFlow (the most naive approach) is the worst.

\subsubsection{Provisioning at Scale}
\label{s:eval_simulation}

\begin{figure*}[t]
\centering
\subfigure[Application load satisfied]{
    \includegraphics[width=0.4\columnwidth,angle=270]{distribution_served.epsi}
    \label{f:distribution_served}
}
\subfigure[Utilization of instances]{
    \includegraphics[width=0.4\columnwidth,angle=270]{distribution_func.epsi}
    \label{f:distribution_func}
}
\subfigure[Inter-rack network utilization]{
    \includegraphics[width=0.4\columnwidth,angle=270]{distribution_inter.epsi}
    \label{f:distribution_inter}
}
\compactcaption{Simulation results to evaluate the efficiency of \Name with a 
 larger deployment}
\label{f:simulation_results}
\end{figure*}

To  examine \Name's ability to efficiently
satisfy application objectives across many \mb chains, we use 
 a custom simulation framework.  Our simulator places
200 chains within a 500-rack data center. The data center is arranged in a
tree topology with 10 VM slots per rack and a capacity of 1Gbps on each
network link. All chains have the same elements and initial instance counts:
workload generators (3 instances), \mb-A (2), \mb-B (1), \mb-C (2), and
servers (4); the capacity of each \mb instance is fixed at 60, 50, and
110Mbps, respectively, and the workload from each generator is 100Mbps. We
perform flow distribution and horizontal scaling (no migration) for each
chain until its full demand is satisfied or no further performance
gain can be achieved.

For ease of visualization and brevity, we only show the results comparing \Name
vs.\ \UniformFlow, noting that the other solution strategies fall in between
these two extremes.

\minisection{Application Objectives} 
We first look at the fraction of each chain's demand that can be satisfied
using \Name and \UniformFlow (\figref{f:distribution_served}). With \Name,
at least 30\% of the demand for all chains is satisfied, with 85\% of demand
satisfied for 40\% of the chains. In contrast, with \UniformFlow, only 20\% of
chains have at least 30\% of their demand satisfied.

\minisection{Efficiency}
Next, we examine how well the \mb instances are utilized.
\figref{f:distribution_func} shows, for each chain, the volume of traffic
served divided by the number of instances used. With \Name, 90\% of instances
process at least 5Mbps of traffic and 50\% process more than 12Mbps; with
\UniformFlow, 90\% of instances process less than 5Mbps of traffic and 50\%
process less than 3Mbps.

Lastly, we examine the amount of inter-rack traffic generated by each chain
(\figref{f:distribution_inter}). Interestingly, with \Name, a higher percent 
of the network is utilized by chains. This is because network-aware flow
distribution allows chains to scale up more and more closely match their
demand, thereby pushing more bytes out into the data center network. On the
whole, the network is more effectively utilized.

\subsection{Scalability of Controllers}
\label{s:eval_scalability}

The primary tasks of \Name' resource controller are monitoring
application performance, computing flow distributions, and placing/migrating
\mbox instances. We can leverage prior techniques for scalable
monitoring~\cite{jain2007star}, so we focus on the resource controller's
ability to perform the later two tasks for large clouds and many tenants. 

We run the resource controller on a machine with an 8-core 2.27GHz CPU and
12GB of RAM. We assume a data center topology consisting of 20K racks, with
20 machines per rack, 625 aggregation switches, and 20 core switches. 
Inter-rack bandwidth and chain traffic volume metrics are randomly generated.
We initially place 1000 tenants with 3 \mb chains each.  Subsequently, we
invoke either flow distribution or placement for many tenants in parallel,
measuring the latency required to complete each task and the maximum
sustained number of operations per second.

\tabref{t:eval_scalability} summarizes our findings. We observe that
computing a flow distribution for a single tenant takes, on average, 701ms.
Our prototype spends a large fraction of this time performing file I/O;
solving the LP using CPLEX takes only 183ms. One controller instance can
compute 51 flow distributions/sec, but we can significantly increase
this capacity by running additional instances of the {\em chain manager}
module (\secref{s:implement}) and assigning subsets of tenants to each
instance, as no synchronization between tenants is required when computing
flow distributions.  For placements, the sustained rate is 67 placements/sec,
each of which takes 506ms on average.  Placement must be coordinated among
tenants, making it inherently less scalable. However, the data center can be
divided into sections, with a separate {\em placement manager} responsible
for each section. Moreover, placement (and migration) is invoked less
frequently than flow distribution, requiring a lower operational capacity.

    
\begin{table}
\centering
\small
\begin{tabular}{l|c|c|c}
    { \bf API }  & { \bf Ops/Sec }  & { \bf Latency } & {\bf Coordinated?}\\
\hline
    Flow distribution & 51 & 183ms & X \\
    Placement  & 67 & 506ms & $\checkmark$ \\
\end{tabular}
\compactcaption{Scalability of primary provisioning tasks}
\label{t:eval_scalability}
\end{table}

\Name' forwarding controller can be scaled using existing
approaches~\cite{koponen2010onix}; we exclude a scalability analysis for
brevity.

\section{Conclusions}

Enterprises today cannot correctly and efficiently deploy virtual middleboxes to improve the
performance and security of cloud-based applications. 
The challenges arise as a
combination of two related factors: (1) the  closed and proprietary nature of
middlebox behaviors (e.g., resource consumption and packet modifications) and
(2) the dynamic and shared nature of the cloud deployments.

%

To address this challenge, we designed a network-aware orchestration layer
called \Name.  \Name allows tenants to realize arbitrarily  complex  logical
topologies by abstracting away the complexity of efficient  \mb composition
and provisioning.  First, to ensure correct  forwarding, even in the presence
of  middlebox mangling and dynamic provisioning, \Name' forwarding controller
combines lightweight software-defined networking mechanisms that also exploits
the virtualized nature of the deployment.  Second,  \Name' provisioning
controller provides a scalable network-aware strategy that synthesizes and
extends  techniques from traffic engineering, elastic scaling, and VM
migration.
 
Using testbed-based live workloads and large-scale simulations, we showed that: (1) \Name ensures efficient and correct composition; (2) \Name' control logic can easily scale to several 
 hundred tenants even on a single server; and (3) \Name generates scalable yet near-optimal provisioning decisions and outperforms a range of strawman solutions illustrating that all the components in \Name' provisioning logic contribute to the overall benefits.

\begin{small}
\bibliographystyle{abbrv}
\bibliography{refs}
\end{small}

\end{document}